\documentclass[sigconf, nonacm]{acmart}
\usepackage{float}
\usepackage{subcaption}
\usepackage{booktabs}
\usepackage{multirow}
\usepackage{enumitem}
\usepackage{makecell} % required for \makecell
\newcommand{\thickhline}{\noalign{\hrule height 1pt}}
\usepackage{epigraph}

\AtBeginDocument{%
  }

\begin{document}

\title{SDR-RDMA: \underline{S}oftware-\underline{D}efined \underline{R}eliability Architecture for Planetary Scale RDMA Communication}


\author{Mikhail Khalilov}
\author{Siyuan Shen}
\author{Marcin Chrapek}
\author{Tiancheng Chen}
\author{Kenji Nakano}

\affiliation{%
 \institution{ETH Zurich}
 \city{Zurich}
 \country{Switzerland}
}

\author{Peter-Jan Gootzen}
\author{Salvatore Di Girolamo}
\author{Rami Nudelman}
\author{Gil Bloch}
\affiliation{%
 \institution{NVIDIA}
 \city{Santa Clara}
 \country{United States of America}
}

\author{Sreevatsa Anantharamu}
\author{Mahmoud Elhaddad}
\author{Jithin Jose}
\author{Abdul Kabbani}
\author{Scott Moe}
\author{Konstantin Taranov}
\author{Zhuolong Yu}
\author{Jie Zhang}
\affiliation{%
 \institution{Microsoft Corporation}
 \city{Redmond}
 \country{United States of America}
}

\author{Nicola Mazzoletti}
\affiliation{%
 \institution{Swiss National Supercomputing Centre (CSCS)}
 \city{Lugano}
 \country{Switzerland}
}

\author{Torsten Hoefler}
\affiliation{%
 \institution{ETH Zurich}
 \institution{Swiss National Supercomputing Centre (CSCS)}
 \city{Zurich}
 \country{Switzerland}
}


\begin{abstract}

RDMA is vital for efficient distributed training across datacenters, but millisecond-scale latencies complicate the design of its reliability layer. We show that depending on long-haul link characteristics, such as drop rate, distance and bandwidth, the widely used Selective Repeat algorithm can be inefficient, warranting alternatives like Erasure Coding. To enable such alternatives on existing hardware, we propose SDR-RDMA, a software-defined reliability stack for RDMA. Its core is a lightweight SDR SDK that extends standard point-to-point RDMA semantics --- fundamental to AI networking stacks --- with a receive buffer bitmap. SDR bitmap enables partial message completion to let applications implement custom reliability schemes tailored to specific deployments, while preserving zero-copy RDMA benefits. By offloading the SDR backend to NVIDIA’s Data Path Accelerator (DPA), we achieve line-rate performance, enabling efficient inter-datacenter communication and advancing reliability innovation for inter-datacenter training.

\end{abstract}


\received{Day Month Year}
\received[revised]{Day Month Year}
\received[accepted]{Day Month Year}

\maketitle

\section{Motivation}

\epigraph{
"We had to scale to more compute, and that compute was not available as part of one cluster. We had to go to multi-cluster training..." --- {\textit{from OpenAI's "Pre-Training GPT-4.5"}\\ \color{blue}{\url{https://www.youtube.com/watch?v=6nJZopACRuQ}}}
}

The scale of a single AI datacenter is constrained by its power plant supply capacity~\cite{semianalysis2024intradcanalysis, daSilva2024google, moseman2024amazon}. As demand for training resources grows, hyperscalers are exploring strategies to utilize the compute capacity of multiple datacenters within a single pre-training job~\cite{team2023gemini, strati2024ml, dong2025beyond}. This necessitates dedicated communication channels between datacenters spanning thousands of kilometers, potentially between continents, such as long-haul black fiber and submarine cables~\cite{balakrishnan2008maelstrom}. Over such distances, a GPU networking stack (e.g., xCCL or MPI~\cite{nccl, gorentlavenkata2025unified, wang2013gpu}), tailored for commodity RDMA Network Interface Cards (NICs), must enable efficient, reliable messaging over lossy connections with millisecond-scale round-trip times (RTTs).

\begin{figure}[t]
  \centering
  \includegraphics[width=\linewidth]{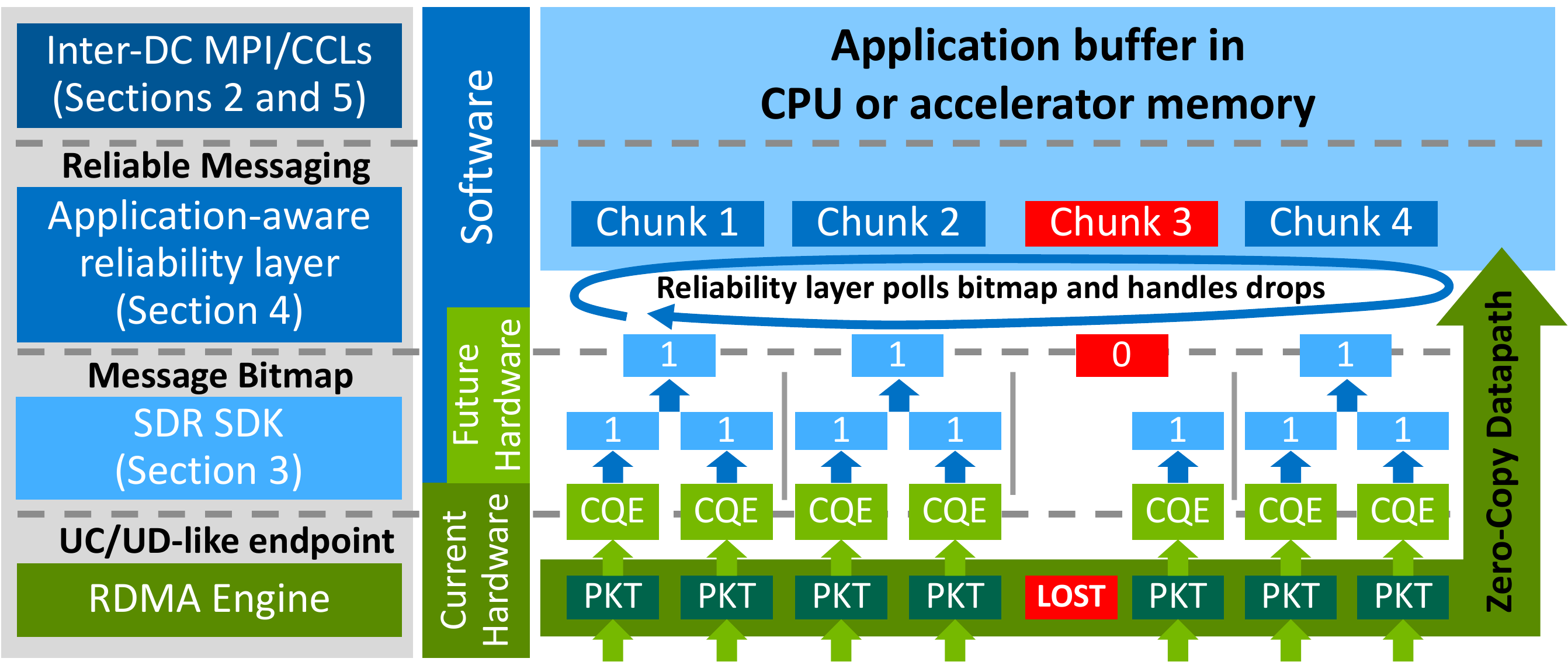}
  \caption{RDMA stack based on the software-defined reliability (SDR) architecture. In all figures throughout the paper, green corresponds to the hardware-related parts of the SDR stack and blue encodes software components.}
\label{fig:urdma-stack}
\end{figure}

Reliability schemes in commodity RDMA NICs (e.g., ConnectX-7/8) are limited to retransmission-based schemes, such as Go-Back-N and Selective Repeat (SR), implemented in the ASIC to support multi-hundred-gigabit bandwidths~\cite{hoefler2023data}. In Section~\ref{sec:inter-dc-challenges}, we illustrate that constraining reliability to retransmission-based schemes can be inefficient for serving traffic through a high-delay, lossy cross-datacenter channel. Our key observation for SR is that message completion time can accumulate multiple RTTs due to retransmission timeouts. SR extensions, such as negative acknowledgment, can minimize the impact of retransmission timeouts in the average case, but not at the tail~\cite{dean2013tail}. A viable alternative could be to avoid relying on acknowledgments and timeouts, and instead leverage erasure coding of application buffers at the transport layer~\cite{zuo2024lowar,zeng2022cutting}.

However, as long as reliability remains part of the NIC ASIC, realizing alternative algorithms in production deployments would take years to materialize, since datacenter operators would need to wait for next-generation silicon to become available~\cite{zahiri2003structured}. A possible solution could be prototyping on FPGAs, but as community experience suggests, FPGAs are generally hard to program and are not supported in commodity datacenter NICs. Furthermore, existing FPGA-based prototypes have not been evaluated for upcoming Tbit/s links~\cite{zuo2024lowar, mittal2018irn, wang2023srnic}.

We solve these problems with a software-defined solution illustrated in Figure~\ref{fig:urdma-stack}. Its bottom software layer --- the software-defined reliability (SDR) middleware SDK --- decouples the low-level details of the packet progress engine from the upper-layer reliability logic. SDR achieves this through novel \textit{partial message completion} API semantics on the receive side. Partial message completion conveys to the reliability algorithm information about message chunks that were dropped in transit, represented as a bitmap.

The key design challenge for SDR is sustaining packet processing at $>100$ Gbit/s with minimal overhead on the CPU and system memory bandwidth. The SDR progress engine uses offloading features in the BlueField-3 SuperNIC to achieve high throughput while exposing a bitmap API to the reliability layer~\cite{nvidiaNVIDIABlueField3Datasheet}. Namely, SDR utilizes hardware-based unreliable RDMA Write for data movement offloading and offloads the packetization and bitmap-updating logic to the Data Path Accelerator (DPA)~\cite{nvidiaDPASubsystem, chen2024demystifying}. These design decisions allow SDR to sustain packet rates on links of up to 3.2~Tbit/s.

Building on the SDR bitmap, we design various SR- and EC-based reliability schemes. We show that the guided choice and performance tuning of an optimal reliability algorithm can improve average and 99.9th percentile RDMA Write completion time by up to $5\times$ and $12\times$, respectively. This is especially critical for cross-datacenter AI collectives, where multi-stage execution causes reliability overheads to compound and degrade end-to-end performance.

\textbf{Our main contributions are:}
\begin{enumerate}
    \item Analysis of inter-datacenter communication challenges.
    \item SDR-RDMA architecture that decouples application-specific reliability logic from low-level packet processing.
    \item SDR-RDMA data path offloading for full line-rate performance on current and next-generation commodity NICs.
    \item Framework to simulate and analyze the performance of SDR-based reliability algorithms in an inter-datacenter setup.
\end{enumerate}

\section{Challenges of inter-DC communication}~\label{sec:inter-dc-challenges}

We show that the NIC reliability algorithm plays a critical role in the performance of point-to-point and collective long-haul communication. We identify two hard requirements for the inter-datacenter networking stack: freedom in the choice of reliability protocol and support for these protocols at line rate in commodity RDMA NICs. In our work, we achieve \textbf{both} with a software-defined solution at the endpoint.

\subsection{There is \textit{no} ideal approach to reliability}

\begin{figure}[H]
    \centering
    \includegraphics[width=\linewidth]{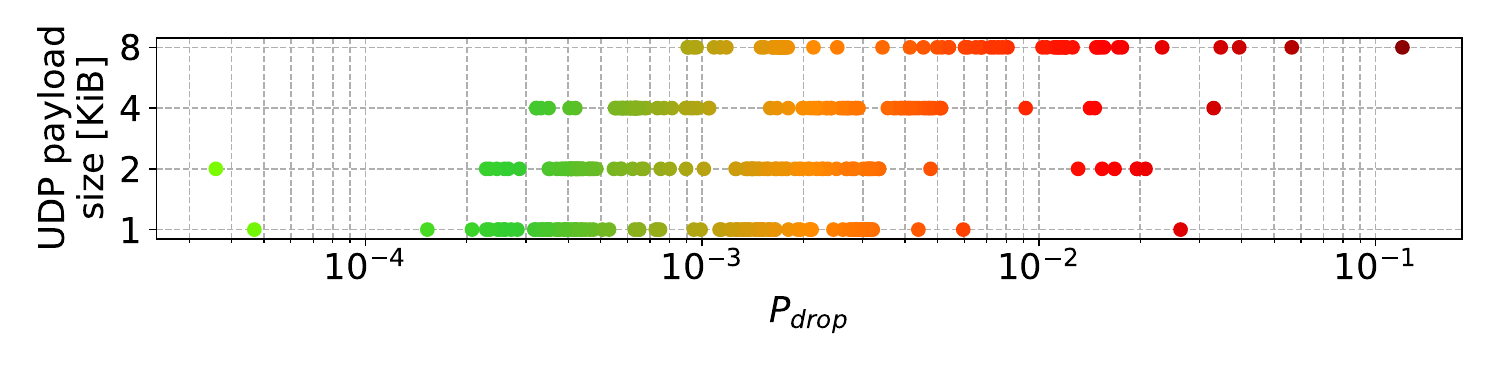}
    \caption{Packet drop rate measured with iperf3 between 16 UDP flows located in Lugano and Lausanne CSCS datacenter sites approximately 350 km apart, connected via a 100 Gbit/s channel. Copper cables are used within both datacenter sites. An optical connection provided by a local ISP is used to connect the two datacenters. For each payload size, drop rates are measured over 200 trials of 15 seconds each, conducted over a 3-day period.}
    \label{fig:drop-rates-cscs}
\end{figure}

In the inter-datacenter scenario, distances, drop rates, and message sizes are not only 2--3 orders of magnitude different from the intra-datacenter case, but can also vary multiple times across deployments. For example, if we consider cross-datacenter setups from Livermore to Oak Ridge and from Lugano to Kajaani, we expect differences of at least 1000~km in total cable lengths, corresponding to approximately $6.5$~ms of added RTT due to geographical features and the proximity of service infrastructure\footnote{We consider paths along the public roads traced by Google Maps between locations of the largest non-private supercomputers in the US (El Capitan, Frontier) and Europe (Alps, Switzerland and LUMI, Finland) from the November 2024 Top 500 list.}.

\begin{figure}[!h]
  \centering
  \includegraphics[width=\linewidth]{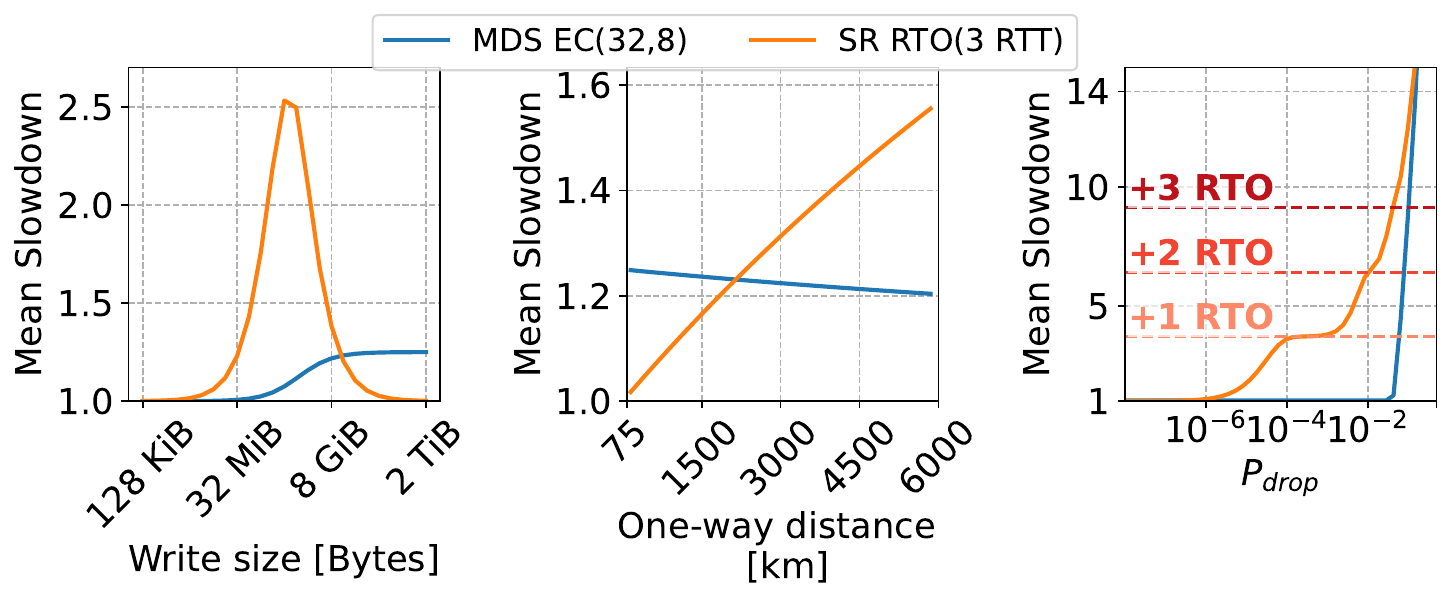}
  \caption{Impact of reliability on message time at 400 Gbit/s. \\[1ex]
\parbox{\linewidth}{\centering
  \begin{tabular}{r@{: }l}
    Left (a)   & 3750 km $=25$ ms RTT, $P_{drop}=10^{-5}$ \\ 
    Middle (b)  & 8 GiB message, $P_{drop}=10^{-5}$ \\ 
    Right (c) & 128 MiB message, 3750 km
  \end{tabular}
}
  }
  \label{fig:motivation}
\end{figure}

Bandwidth and drop rates in inter-datacenter links can vary widely due to factors like cable technology, budget, and QoS policies. Figure~\ref{fig:drop-rates-cscs} shows UDP drop rates between two datacenters connected via an optical link from a publicly funded ISP. On both sides, traffic was isolated from other intra-datacenter traffic, and trials with endpoint-side drops --- tracked through Linux kernel counters --- were excluded. We observe up to three orders of magnitude variation in drop rates across trials for the same packet size, with drop rates increasing for larger packets: from $10^{-4}$ to $10^{-2}$ for 1KiB, and from $10^{-3}$ to over $10^{-1}$ for 8~KiB packet sizes, respectively. Such variability and its correlation with packet size suggest significant switch buffer congestion on the ISP side. In contrast, data collected across 15 Microsoft datacenters show that pre-Ethernet-FEC drop rates on private optical networks can be as low as $10^{-8}$~\cite{zhuo2017understanding}.

Figure~\ref{fig:motivation} shows the impact of long-haul channel parameters on reliability performance at 400~Gbit/s. We compare two schemes (see Section~\ref{sec:reliability-layer}): a standard Selective Repeat (SR) protocol and an Erasure Coding (EC) scheme. Unlike SR, EC does not rely on retransmissions but instead sends parity data with the original message. While EC is not supported by commodity NICs, recent FPGA-based results at 10~Gbit/s\cite{zuo2024lowar} suggest that it could reduce retransmission overhead in high-delay, lossy channels.

We observe that there is no single "best" reliability scheme. In Figure~\ref{fig:motivation}a, SR reaches peak slowdown at the point when at least one packet in the message is likely to be dropped, given a drop rate of $10^{-5}$. With SR, the retransmission cost for this drop cannot be hidden in the pipeline --- each retransmission significantly contributes to the message completion time. In contrast, the EC-based scheme remains close to ideal performance because the receiver can recover from drops in-place using parity data. Above $2^{35} = 32$~GiB, the message becomes large enough for packet injection time to dominate over channel delay. For such a "large" message, while SR can efficiently hide the latency of retransmissions behind the time required to inject the message, the EC-based scheme consumes a portion of the channel bandwidth to send parity data.

In Figure~\ref{fig:motivation}c, with a "small" 128~MiB message at drop rates above $10^{-4}$, we observe an increase in completion time from $3\times$ to $10\times$ due to a single packet requiring multiple retransmission rounds. In Figure\ref{fig:motivation}b, as the cable distance between datacenters increases, an 8~GiB message that was previously considered "large" (dominated by the injection time) becomes "small" (dominated by RTT delay). In the "large" interval, SR outperforms EC, whereas in the "small" interval, the trend is reversed.

The above result suggests that supporting various reliability schemes in the NIC is essential to enable efficient communication between two datacenters. Furthermore, when considering a scenario with more than two datacenters, a single endpoint might communicate with remote endpoints at varying distances. Achieving optimal message completion times in this scenario may require per-connection reliability protocol provisioning.

\begin{table*}
\centering
\begin{tabular}{cll}
\textbf{Subset}                              & \textbf{API call}                                                                          & \textbf{Description}                                                                       \\ \thickhline
                                             & {\color[HTML]{000080} ctx *context\_create(char *dev\_name, dev\_attr *dev\_attr);}        & Allocate HW resources (CQs, DPA threads) shared by QPs.              \\
                                             & {\color[HTML]{000080} qp *qp\_create(ctx *ctx, qp\_attr *qp\_attr);}                       & Create a queue pair within a context.                                      \\
                                             & {\color[HTML]{000080} int qp\_info\_get(qp *qp, void *info);}                              & Retrieve QP info for out-of-band exchange.                                \\
\multirow{-4}{*}{\makecell{Data path\\setup}} & {\color[HTML]{000080} int qp\_connect(qp *qp, void *remote\_qp\_info);}                    & Establish connection between queue pairs using QP info.                  \\ \hline
Memory                                       & {\color[HTML]{000080} mr *mr\_reg(ctx *ctx, void *addr, size length, flags flags);}        & Register memory for send/receive via QPs in the context.                          \\ \hline
                                             & {\color[HTML]{000080} int send\_stream\_start(qp *qp, start\_wr *wr, snd\_handle **hdl);}  & Create streaming send (w. Imm) message context. \\
                                             & {\color[HTML]{000080} int send\_stream\_continue(snd\_handle *hdl, continue\_wr *wr);}     & Send new chunk(s) into a remote buffer (stream) by an offset.                  \\
                                             & {\color[HTML]{000080} int send\_stream\_end(snd\_handle *hdl);}                            & Indicate that no new chunks will be added to the stream. \\
                                             & {\color[HTML]{000080} int send\_post(qp *qp, snd\_wr *wr, snd\_handle **hdl);}             & Initiate a one-shot send (w. Imm) message. \\
\multirow{-5}{*}{\makecell{Send}}            & {\color[HTML]{000080} int send\_poll(snd\_handle *hdl);}                                   & Poll for a send message completion.        \\ \hline       
                                             & {\color[HTML]{000080} int recv\_post(qp *qp, rcv\_wr *wr, rcv\_handle **hdl);}             & Post a receive message buffer.                                                           \\
                                             & {\color[HTML]{000080} int recv\_bitmap\_get(rcv\_handle *hdl, uint8 **bitmap, size *len);} & Get pointer to the bitmap associated with a receive buffer.                      \\
                                             & {\color[HTML]{000080} int recv\_imm\_get(rcv\_handle *hdl, uint32 *immediate);}            & Retrieve immediate data if it is ready.                              \\
\multirow{-4}{*}{Receive}                    & {\color[HTML]{000080} int recv\_complete(rcv\_handle *hdl);}                               & Marks a receive message as complete.                        
\end{tabular}
\caption{SDR API overview. Object metadata (e.g., Write size and offset) is encapsulated into the C structs.}
\label{tbl:urdma-api}
\end{table*}

\subsection{Need for a software-defined solution}

From the bottom-level protocol perspective, current generation NICs (e.g., RoCEv2) can serve inter-datacenter traffic out-of-the-box. Their physical, link, and network layers are compatible with Ethernet and Internet protocols and can operate in lossy mode~\cite{winkler2015scinet}. The custom RDMA transport logic (e.g., congestion control and reliability) is implemented above the commodity protocols and runs at the endpoints (e.g., QPs).

We begin by analyzing the Reliable Connection (RC) Verbs transport, which provides the reliable Write primitive used by distributed training libraries~\cite{gorentlavenkata2025unified, nccl, wang2013gpu}. To our knowledge, current RDMA NICs from Broadcom, Intel, Microsoft, and NVIDIA support only retransmission-based RC protocols, typically implemented in the NIC's ASIC for performance~\cite{hoefler2023data}. Since ASIC development cycles span 3–4 years, adopting and optimizing new reliability protocols, such as erasure coding (EC), would take years to materialize.

We believe a software-defined approach to RDMA reliability is the right path forward. Just as QUIC enabled rapid innovation in transport protocols on top of UDP~\cite{langley2017quic, marty2019snap}, a software layer over unreliable RDMA transports can foster fast-paced development, adoption, and optimization of application-aware reliability protocols tailored for datacenter environments.

\subsection{Transport design challenges}\label{sec:design-challenges}

Existing RDMA implementations offer support for unreliable transports, similar to how the operating system provides UDP service to QUIC. For example, the Verbs API supports Unreliable Datagram (UD) and Unreliable Connected (UC)~\cite{ibspec}. Libfabric API~\cite{ofi-libfabric} also offers datagram messaging (DGRAM-MSG) and datagram RMA (DGRAM-RMA) as equivalents. Without loss of generality, we focus our discussion on Verbs:
\begin{itemize}[leftmargin=*]
\item \textit{UD} offers a two-sided per-packet service. Emulation of reliable Write semantics on top of UD transport is feasible. However, due to the possibility of out-of-order packets (e.g., because of drops) it comes at the cost of intermediate packet staging in the host CPU or NIC memory on the receive side~\cite{khalilov2024network, li2023flor}.
\item \textit{UC} offers unreliable multi-packet Writes. With UC, out-of-order packets are not problematic, because the sender side determines the target memory (address) of Write. If at least one packet within the UC message is dropped, the whole message will be dropped.
\end{itemize}

UC is ideal for building zero-copy transport, but its coarse completion semantics are unsuitable for reliability layers. For instance, if one 4~KiB packet is lost in a 1~GiB Write, the NIC considers the entire Write lost—forcing the application to retransmit all 1~GiB, wasting time and bandwidth.

To address this, we introduce a lightweight middleware between UC and the reliability layer that delivers messages in \textit{chunks}, each aligned with the MTU. A \textit{bitmap} tracks received chunks, allowing the reliability layer to process available data. EC can use this to identify and repair losses with parity, while SR uses it to report dropped chunks to the sender.

\section{SDR middleware}

Our goal is to enable innovation of reliability algorithms in current generation RDMA NICs. We achieve this goal with the software-defined reliability (SDR) SDK, a middleware that extends conventional RDMA completion semantics to support unreliable arbitrary-length messaging with a partial completion bitmap. The bitmap can be used by the reliability layer to locate drops within a message.

\subsection{Partial message completion API}

Table~\ref{tbl:urdma-api} presents the SDR API. We discuss its novel features.

\subsubsection{Partial completion bitmap} A key feature of SDR is a partial completion bitmap enabled on top of standard unreliable RDMA transports (e.g., unreliable Write). The bitmap is a lightweight software abstraction that decouples the reliability protocol logic running above it from the RDMA progress engine running below it. SDR users (e.g., a reliability layer) can post a receive buffer and track chunks that have been received by polling the bits in the associated bitmap, while packet (de)fragmentation progress is offloaded to the NIC (handled by the SDR runtime behind the scenes).

A single bit in the bitmap corresponds to a message \textbf{chunk}—a contiguous block of bytes within a receive buffer. Chunk size is a multiple of the network Maximum Transfer Unit (MTU) and is configurable by the user. It can be tailored to the specific needs of the application running on top of SDR. For example, the bitmap resolution can be chosen to mask drop bursts within the same chunk; with a chunk size of 16 packets, dropping 7 packets inside a chunk would appear to the upper layer as a single chunk drop.

\subsubsection{Sender-side optimizations} SDR supports two send types:
\begin{itemize}[leftmargin=*]
    \item \textit{Streaming} send offers fine-grained control: new chunk(s) can target any offset in the remote buffer and are added to the send stream. For reliability, a typical use case is retransmission --- for example, resending a chunk after a timeout.
    \item \textit{One-shot} send prioritizes efficiency when transmitting large contiguous data blocks in a single operation. Once all chunks are injected, the message context is destroyed.
\end{itemize}

Both APIs are asynchronous to enable overlap between computation and network injection. By providing two distinct primitives we allow the sender to select the appropriate granularity, while the SDR backend can optimize the two send paths independently.

\subsubsection{Order-based message matching} SDR is a message-based API: on the receiver side, there is an association between the receive buffer and a bitmap state, while the streaming API assumes that new chunks are added to the same stream queue associated with a fixed remote buffer. In SDR, message matching between sender and receiver is \textit{order-based}: the sender’s send messages "land" in the receiver's buffers in the order they were posted.

For example, let's assume that the receiver posts two receive buffers in sequence: Recv1, Recv2. When the sender posts the send messages (e.g., Send1, Send2), the order-based matching ensures that Send1 targets Recv1, and Send2 targets Recv2. Because SDR matching is order-based, explicit buffer metadata (e.g., remote memory key, virtual address, etc.) does not need to be exchanged between receiver and sender, apart from ensuring that the receive is posted before the corresponding send is issued.

\subsection{Messaging protocol}

\begin{figure}[h]
  \centering
  \includegraphics[width=\linewidth]{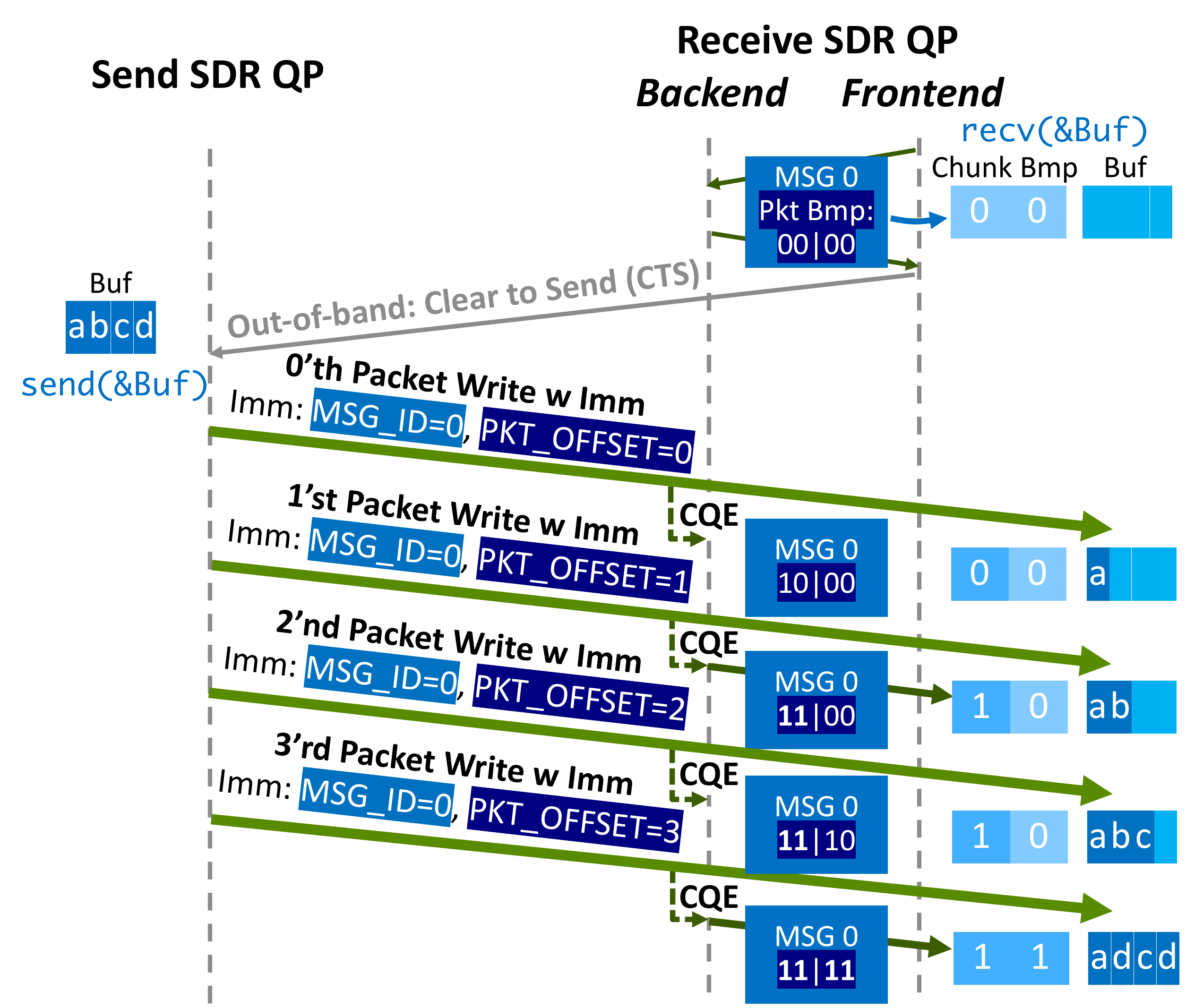}
  \caption{Example of SDR one-shot send data path with four packets, two packets per receive bitmap chunk. The protocol backend can be offloaded to the NIC.}
  \label{fig:urdma-protocol}
\end{figure}

We illustrate the internal SDR messaging protocol with the client-server example in Figure~\ref{fig:urdma-protocol}.

\subsubsection{Backend/frontend design} In the backend, SDR relies on UC as a zero-copy engine for chunk delivery (UD can still be used with the limitations outlined in Section~\ref{sec:design-challenges}).

While sending each chunk with a single RDMA write-with-immediate over UC would be the simplest solution, this could result in the entire chunks being dropped if packets are reordered by ISPs.

This limitation arises from the behavior of the receiver’s UC QP when handling multi-packet messages. Specifically, the UC QP maintains an expected packet sequence number (ePSN)~\cite{ibspec} that is incremented with each received packet and resets at the start of every new message. If an incoming packet PSN does not match the ePSN, the entire message is dropped due to PSN mismatch.

To overcome this limitation, we instead issue one RDMA Write-with-immediate operation per packet, making each packet being handled as a single message. While this strategy increases the backend processing load, as it requires tracking each individual packet separately, it enables SDR to handle out-of-order packet delivery.

To efficiently handle per-packet tracking and transport-level queue management without increasing CPU load, we design the internal SDR messaging protocol with offloading in mind. The backend maintains a per-packet bitmap for each message, which is \textit{coalesced} into a frontend chunk bitmap. A chunk is only signaled when all its packets arrive. Through this decoupling, we enable backend offloading to the Data Path Accelerator (DPA), a programmable engine in the BlueField-3 SuperNIC designed for parallel traffic processing~\cite{nvidiaDPASubsystem, nvidiaNVIDIABlueField3Datasheet, chen2024demystifying}.

\subsubsection{QP creation and connection establishment}  
\begin{figure}[h]
  \centering
  \includegraphics[width=\linewidth]{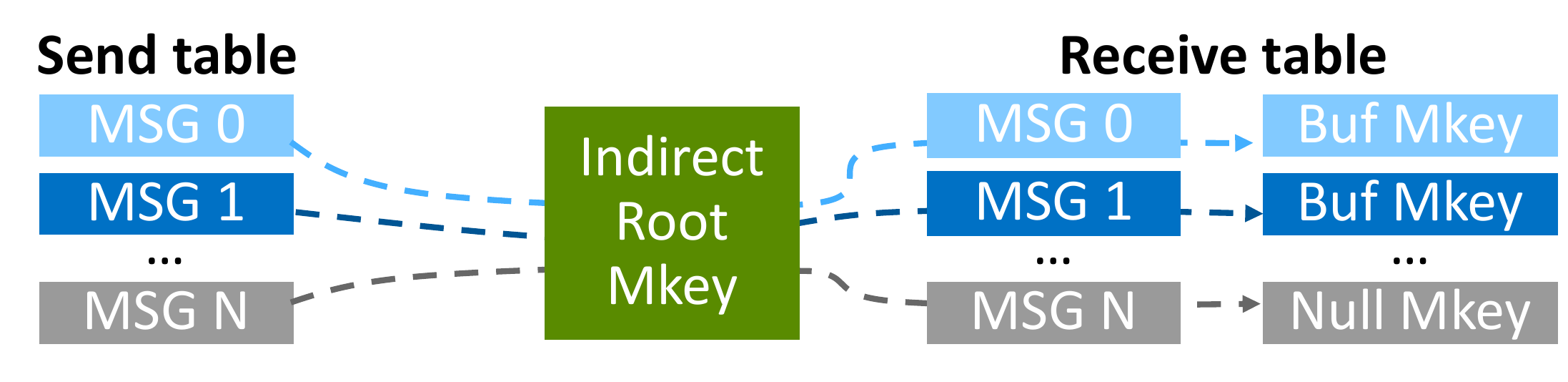}
  \caption{Sender and receiver SDR QPs share a zero-based root memory key, enabling offset-based addressing of receive buffers. For a QP with maximum message size $M$, message $i$ targets the offset range $[i \cdot M, i \cdot M + M)$.}
  \label{fig:urdma-msg-table}
\end{figure}

During QP setup, send and receive backends populate message tables and associate them with each other via an indirect memory key exchanged out-of-band (Figure~\ref{fig:urdma-msg-table}). The runtime allocates internal buffers for per-packet (backend) and chunk (frontend) bitmaps, based on the user-defined maximum message size and bitmap chunk size (4-packet messages with 2-packet chunks in Figure~\ref{fig:urdma-protocol}).

\subsubsection{Posting buffers}
During the post-receive call, the backend allocates a context slot in the message table, updates the indirect root memory key table with the user buffer’s key, and sets up per-packet and chunk bitmaps. After posting the receive buffer, the receiver sends an out-of-band clear-to-send (CTS) signal. The sender can then begin writing new data into the buffer.

\subsubsection{Data path}

The SDR send operation results in a sequence of single-packet unreliable Writes --- four Writes in our example target the receiver's root indirect memory key, where the offsets $0$, $MTU$, $2MTU$, and $3MTU$ are backed by the receive buffer.

The NIC’s RDMA engine writes the UC packet payload directly into user buffer. The receive backend then obtains a completion (CQE or cookie) for the packet. Each CQE includes 32-bit \textit{transport immediate data}, divided into three fields:
\begin{enumerate}[leftmargin=*]
    \item 10 bits for the message ID (light blue in \textit{Imm} of Figure~\ref{fig:urdma-protocol}), allowing up to 1024 in-flight message descriptors per QP.
    \item 18 bits for the packet offset (dark blue in \textit{Imm} of Figure~\ref{fig:urdma-protocol}), supporting up to 1~GiB message size with a 4~KiB MTU.
    \item 4 bits for user immediate reconstruction (not shown); for messages with user immediate, the sender backend samples fragments of it into this field.
\end{enumerate}
This $10+18+4$ bit split reflects our use-case needs. Alternative splits, such as $8+22+2$, can be used to support larger messages.

Using the immediate fields, the receive backend locates the message descriptor and computes the bitmap offset for the packet. In Figure~\ref{fig:urdma-protocol}, a packet with offset zero sets the first bit in the packet bitmap; the second packet sets the second bit. Once all bits for the first buffer chunk are set, the backend updates the corresponding bit in the frontend’s chunk bitmap. The same procedure is applied to the next two packets.

\subsection{Late packet arrival protection}

\begin{figure}[h]
  \centering
  \includegraphics[width=\linewidth]{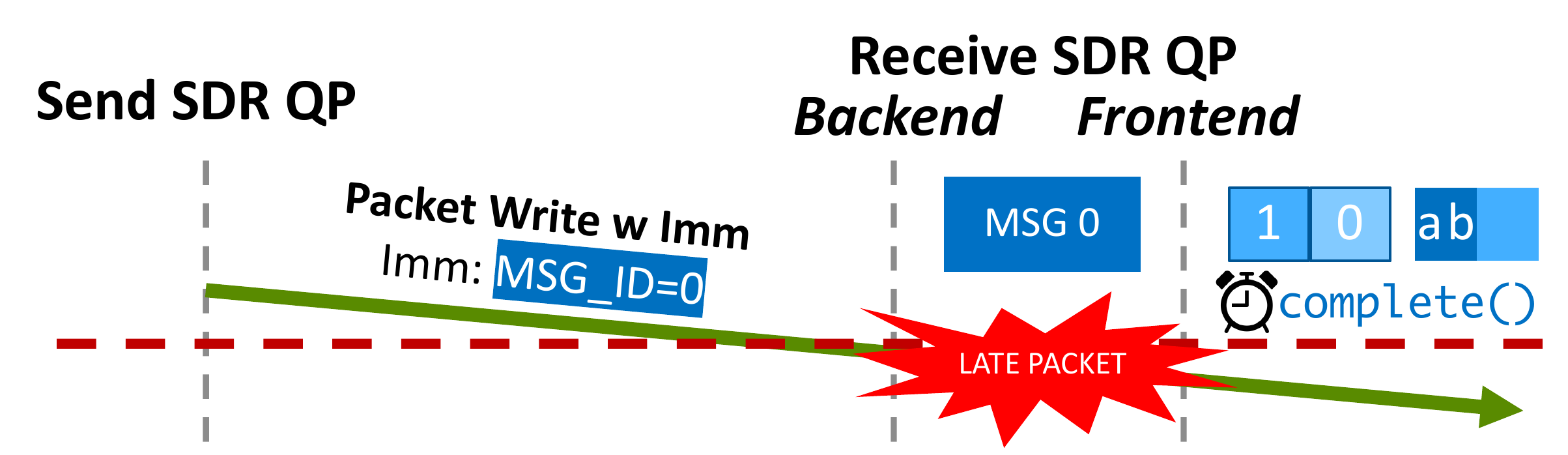}
  \caption{Late packet arrival problem.}
  \label{fig:late-packet-problem}
\end{figure}

\subsubsection{Early receive completion}  
In the example in Figure~\ref{fig:urdma-protocol}, the application may complete a posted receive even if some chunks have not yet arrived. For instance, a receiver-side timeout (Figure~\ref{fig:late-packet-problem}) can trigger early completion while packets are still in flight. Because the data path is one-sided, the sender remains unaware and may continue injecting packets for a message that no longer exists. In both cases, receive context resources (buffer and bitmaps) must be protected from late arrivals.

\subsubsection{Message ID space wraparound}  
Encoding the message ID in the transport immediate field is limited to 10 bits, causing wraparound every 1024 messages. If this occurs too fast, late packets may corrupt the buffer and bitmaps of a newly posted message reusing the same ID. At 800~Gbit/s and 16~MiB messages, wraparound occurs about in 100~ms --- safe for RTTs below 100~ms. However, switch buffering, faster links, or smaller messages reduce this safety margin.

We solve both problems with a two-stage protection:

\begin{enumerate}[leftmargin=*]
\item When a message is completed, the corresponding entry in the root indirect memory key table is updated to point to a special NULL memory key that discards packet payloads (via \texttt{ibv\_alloc\_null\_mr()} in Verbs). Writes to the NULL memory key generate completion entries for late packets, which are filtered at the second stage.
\item To prevent bitmap corruption by late packet completion entries, we introduce the concept of \textit{message generations}. Upon its creation, an SDR QP allocates multiple internal QPs, each associated with its own generation—for example, 4 internal QPs to support 4 message ID generations. The backend tracks the current generation for each message ID slot. For each completion entry, the current message slot generation is checked against the generation of the QP that delivered the entry. If they do not match, the completion entry is discarded.
\end{enumerate}

The generation mechanism increases tolerance for late packets. Although it requires extra QPs~\cite{wang2023srnic}, their sequential use is enabled by traffic’s temporal locality and the rarity of late packet events.

\subsection{Backend acceleration}

\begin{figure}[h]
  \centering
  \includegraphics[width=\linewidth]{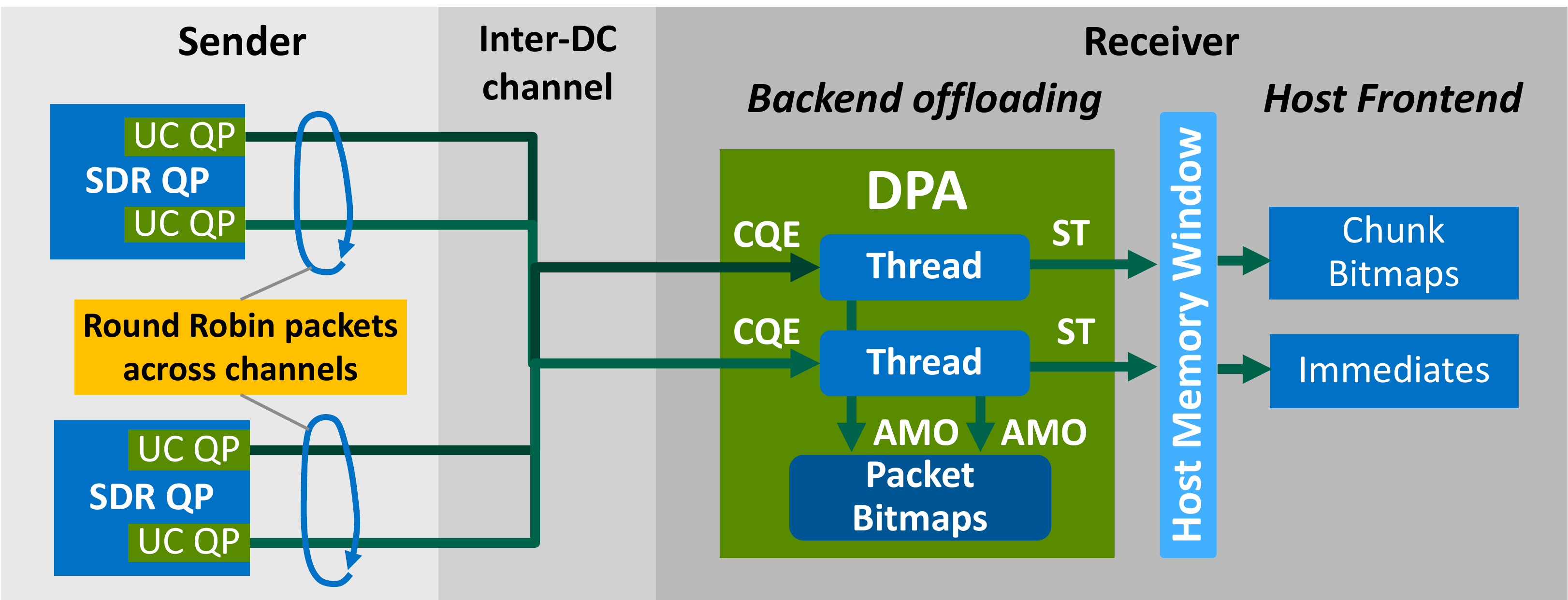}
  \caption{Multi-channel SDR offloading architecture.
}
\label{fig:multi-channel-offloading}
\end{figure}

Packet processing at line rates above 100~Gbit/s requires endpoint parallelism\cite{belay2014ix, hoefler2017spin, khalilov2024network}. To this end, the SDR backend offloads packetization and bitmap processing to the Data Path Accelerator (DPA), a programmable component of the BlueField-3 and ConnectX-8 SuperNICs~\cite{nvidiaDPASubsystem, nvidiaNVIDIABlueField3Datasheet, nvidiaConnectX8SuperNICDatasheet, chen2024demystifying}. The DPA, with its 256 energy-efficient hardware threads, is well-suited for parallel processing of SDR transport Write completions. We implement send- and receive-side offloading using the user-space Flex IO SDK API, part of the DOCA SDK. We focus on receive-side offloading, which handles the major of SDR’s data path logic.

\subsubsection{Multi-channel design} We extract traffic parallelism using the multi-channel design shown in Figure~\ref{fig:multi-channel-offloading}. For each message generation, the backend allocates multiple transport QPs, which serve traffic in parallel and act as independent channels. The sender distributes packets of a message across channels, allowing packets from different channels to be processed concurrently on the DPA. This is achieved by mapping different channels to separate completion queues, each polled by a different receive DPA worker thread. The multi-channel design enables linear scaling of protocol bandwidth with the number of DPA worker threads. Further, by spreading traffic across channel QPs, SDR could leverage intra-datacenter multi-pathing (e.g., ECMP~\cite{ecmp, gangidi2024rdma}) and multi-plane networks~\cite{nvidiaSpectrumXDatasheet,an2024fire}.

\subsubsection{Receive DPA worker} For each packet completion, the DPA worker thread validates the packet generation and initiates bitmap processing. The worker uses the packet offset from the transport immediate to locate and atomically update the corresponding chunk in the per-packet bitmap stored in DPA memory. The worker thread that receives the final missing packet within a chunk also updates the host-side chunk bitmap over PCIe.

\section{Example reliability layers}\label{sec:reliability-layer}

We discuss how the SDR messaging protocol with bitmap support can be used as a base primitive to express reliable RDMA Write. We compare two example strategies for end-to-end error correction on top of SDR: Selective Repeat (SR) from the Automatic Repeat reQuest (ARQ) family of protocols, and Erasure Coding (EC) from the Forward Error Correction (FEC) family. We choose SR as an example ARQ scheme since it can be proven theoretically that SR efficiency is at least as good as Go-back-N's~\cite{bertsekas2021data}.

In ARQ, chunk error correction requires at least one RTT to trigger retransmission, while FEC uses extra bandwidth to speculatively send parity chunks. We develop a theoretical framework for message completion time and use it to analyze how the core bandwidth-latency trade-off emerges in RDMA Write over a lossy, delayed channel.

\subsection{Protocols}

\begin{figure}[h]
  \centering
  \includegraphics[width=\linewidth]{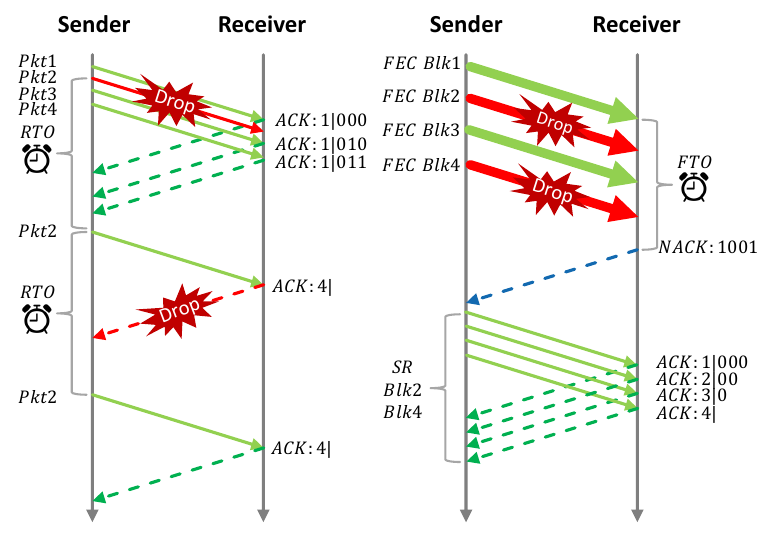}
  \caption{SDR-based SR and EC reliability protocols. Left: a 4-packet message delivered with SR. Right: an 8-packet message delivered using $EC(2,1)$.}
  \label{fig:reliability-protocols}
\end{figure}

We illustrate both protocols in Figure~\ref{fig:reliability-protocols}. We assume that the client and server have established two uni-directional connections:
\begin{itemize}[leftmargin=*]
\item \textit{Data-path SDR QP:} for zero-copy data transfer.
\item \textit{Control-path UC (or UD) QP:} to exchange protocol acknowledgment packets (ACKs) with low overhead.
\end{itemize}

Notice that the two-connection design is not a hard requirement, and we choose it solely for the purposes of our analysis. The SDR middleware API leaves the control path wireup logic to the application implementing reliability, thereby enabling application-aware optimizations such as the optimized rendezvous protocol~\cite{sur2006rdma}.

\subsubsection{SR-based reliability} We illustrate a timeout-based scheme inspired by TCP selective acknowledgment~\cite{mathis1996rfc2018}, treating it as the most general design. More advanced schemes, such as negative acknowledgment (NACK)~\cite{lu2018multi} and retransmission timeout (RTO) tuning~\cite{sarolahti2003f}, can also be supported. We analyze how NACK could improve SR performance in Section~\ref{sec:two-dcs-case-study}.

\textit{SR sender} uses streaming SDR send to inject message chunks into the network. Each chunk is assigned a timeout ($RTO = RTT + \alpha \cdot RTT$), where $\alpha$ reflects switch buffering along the sender–receiver path. When the $RTO$ expires, the sender retransmits the chunk. Upon receiving an ACK, the sender removes all chunks within the acknowledged range from the retransmission queue.

\textit{SR receiver} periodically polls the message chunk bitmap and sends ACKs to the sender. Each ACK compactly encodes the receiver’s bitmap in two parts:
\begin{itemize}[leftmargin=*]
\item \textit{Cumulative ACK:} the highest chunk sequence number for which all previous chunks have been received.
\item \textit{Selective ACK:} a portion of the bitmap (as much as fits in the ACK payload), starting from the cumulative ACK.
\end{itemize}

\subsubsection{EC-based reliability}
With the EC-based scheme, costly chunk retransmissions can be avoided by computing (EC encoding) additional parity chunks for the data chunks. Parity chunks are sent speculatively alongside the data chunks—in case any data chunks are missing and enough parity chunks are received, the missing data chunks can be recovered (EC decoding).

\textit{EC sender} splits the message into $L$ data submessages of $k$ chunks each, where \(L = M / k\). Each submessage is erasure coded with $m$ parity chunks to form a corresponding parity submessage, resulting in $2L$ one-shot SDR sends. Encoding can proceed asynchronously (e.g., on spare CPU cores~\cite{chrapek2023hear}) during data transmission, leveraging SDR’s non-blocking send semantics. After injecting all submessages and receiving a positive acknowledgment, the sender releases the message buffer.

\textit{EC receiver} polls the bitmap and sends a positive ACK once enough chunks are available to recover all data submessages. If any submessage bitmap indicates missing chunks, the receiver supplies the received data and parity chunks to the EC decoder.

\textit{Fallback scheme:} We must consider an edge case where the receiver cannot recover a data submessage. For example, in Maximum Distance Separable (MDS) codes, when the total number of dropped chunks across the data and corresponding parity submessage exceeds $m$. Our fallback strategy is to switch to Selective Repeat for the failed data submessages. When the receiver sees the first chunk of a message (the first bit in the bitmap is observed), it sets a fallback timeout ($FTO = M / BW_{channel} + 0.5 \cdot \alpha \cdot RTT$). We halve the $\alpha$ coefficient from the SR scheme, as only half of the buffering along the $RTT$ path needs to be accounted for. Upon $FTO$ expiration, the receiver sends a negative ACK (NACK) listing the failed data submessages. A global timeout is also set at message posting to prevent deadlock, though it is assumed to be rarely triggered.

Higher parity-to-data ratios in EC algorithms improve tolerance to chunk drops but increase channel bandwidth usage. We build mathematical intuition for this trade-off in the next subsection and evaluate recovery behavior across EC configurations in Section~\ref{sec:two-dcs-case-study}.

\subsection{Message completion time model}~\label{sec:fct-model}

We demonstrated SR- and EC-based reliability schemes built on SDR and now introduce a statistical framework to evaluate their performance. Our model captures key inter-datacenter parameters: drop rate, delay, bandwidth, and application message size. It is released as an open-source Python library, enabling system architects to design and tune the reliability layer to specific RDMA deployments.

\subsubsection{Mathematical notation}
Our model notation is as follows:
\begin{itemize}[leftmargin=*]
\item $M$ is the message size in chunks of receive-side bitmap.
\item $T_{protocol}(M)$ is the Write completion time at the sender side when the reliablity $protocol$ is used (i.e., the time interval between the injection of the first chunk and the ACK reception for the last unacknowledged chunk).
\item $T_{INJ}$ is the time to inject a chunk into the network (the inverse of chunk size divided by link bandwidth~\cite{alexandrov1995loggp}).
\item $P_{drop}$ is the probability of a chunk drop on the sender–receiver path. We assume that $P_{drop}$ is i.i.d for each chunk.
\end{itemize}

\subsubsection{Selective Repeat}

Message completion time for SR must account for all possible positions of a chunk drop.

For the \(i\)-th chunk (with \(i = 1, \dots, M\)), we define its time:
\[
X_i = t_{\text{start}}(i) + O\,\bigl(Y_i - 1\bigr),
\]
where:
\begin{itemize}[leftmargin=*]
    \item \(t_{\text{start}}(i)=i \cdot T_{INJ}\) is the start time for the \(i\)-th chunk,
    \item \(O=RTO+T_{INJ}>0\) is the overhead incurred for each drop,
    \item \(Y_i\) is a geometric random variable with success probability \(1 - P_{drop}\). \(Y_i\) gives a lower bound on the number of transmissions needed for successful delivery of the \(i\)-th chunk.
\end{itemize}

We define the overall completion time of a message as the \textit{maximum} over all individual chunk times:
\[
T_{SR}(M) \ge \max_{1 \le i \le M} X_i + RTT.
\]

Two scenarios must be distinguished in this formulation:

\begin{enumerate}[leftmargin=*]
    \item \(t_{\text{start}}(M) \le RTO\): In this case, the initial offset \(t_{\text{start}}(i)\) ensures that all retransmitted chunks are reinjected into the network at different times from the initial \(T_0\). For example, if chunks \(i\) and \(i+1\) are dropped on their first transmissions, their retransmission timeouts will differ by \(T_{INJ}\), and they will be reinjected at times \(T_0 + RTO_i\) and \(T_0 + RTO_i + T_{INJ}\), respectively.
      
    \item \(t_{\text{start}}(M) > RTO\): In this "large" message case, our derivation for \(E[T_{SR}]\) becomes a \textit{lower bound} on the expected message completion time. Consider a scenario where the first chunk is dropped. Its retransmission timestamp occurs \textit{before} the last chunk is injected into the network for the first time at \(t_{\text{start}}(M)\), thereby violating serialization.
\end{enumerate}

Two methods are employed to evaluate \(T_{SR}(M)\): a stochastic simulation and analytical solution for its expected value. We provide the derivation of the analytical solution in Appendix~\ref{sec:fct-derivation}.

\subsubsection{EC-based reliability}
Recall that for an erasure code with \(k\) and \(m\) chunks in data and parity submessages, respectively, the total number of independently erasure-coded data submessages is \(L = M / k\). Let's denote the probability of successful recovery of a data submessage (which is a function of \(P_{drop}\)) as \(P_{EC(k,m)}\).

The expected number of failed data submessages is
\[
E[\text{failures}] = L \cdot \left(1 - P_{EC(k,m)}\right),
\]
and the probability of at least one data submessage failing (i.e., the probability of fallback to Selective Repeat) is
\[
P_{fallback}^{EC} = 1 - \left(P_{EC(k,m)}\right)^L.
\]

Let the parity ratio be \(R = k / m\) for \(EC(k, m)\). In our protocol, the receiver sets an SR fallback timeout (\(FTO\)) as soon as any chunk of the message is received (the first bit is observed in the bitmap of any submessage):
\[
FTO = (M + \lceil M / R \rceil) T_{INJ} + \beta \cdot RTT.
\]

Assuming no network congestion or resource contention, two scenarios may occur on the receiver after \(FTO\) is set:
\begin{enumerate}[leftmargin=*]
    \item By the time \(FTO - \beta \cdot RTT\), the receiver has received enough chunks to recover all data submessages and sends back an ACK.
    \item \(FTO\) expires, and the receiver sends an EC NACK to the sender, requesting selective repeat of the failed data submessages (each consisting of \(k\) data chunks).
\end{enumerate}

Assuming full overlap of data injection and parity computation, the lower bound on \(E[T_{EC}(M)]\), is a sum of the following terms:
\begin{enumerate}[leftmargin=*]
\item Base time to initially send data and parity chunks:
\[
(M + \lceil M / R \rceil) \cdot T_{INJ},
\]
\item \textit{Plus} the expected time spent in timeout and EC NACK delivery:
\[
P_{fallback}^{EC} \cdot (RTT + \beta \cdot RTT),
\]
\item \textit{Plus} the expected time to retransmit the failed submessages:
\[
E\left[T_{SR}(E[\text{failures}] \cdot k)\right].
\]
\end{enumerate}

\section{Evaluation}~\label{sec:eval}

We examine the following research questions:
\begin{enumerate}[leftmargin=*]
\item How do the discussed SDR-based reliability algorithms perform in various long-haul scenarios?
\item How does the choice of reliability scheme impact performance of inter-datacenter AI training traffic?
\item Can the SDR middleware enable partial message completion service at line rate on the current and next-generation NICs?
\end{enumerate}

\subsection{Experimental setup}~\label{sec:experimental-setup}

\subsubsection{SDR simulation setup}

We implement, using Python~3, a stochastic simulation for the Write completion time with the SR and EC SDR protocols presented in Section~\ref{sec:reliability-layer}. We validate simulation results against the analytical expectation for message completion time. The mean of 1000 samples from the stochastic model matches the analytical solution within 5\% accuracy.

We study two scenarios for the SR-based algorithm. In the first, SR RTO, we set the SR chunk timeout to 3 network RTTs. In the second scenario, SR NACK, we reduce the RTO to 1 network RTT as a best-case approximation of the negative acknowledgment (NACK) optimization. In SR NACK, the receiver sends a negative signal to the sender indicating the specific location of the dropped chunk; therefore, the sender can initiate retransmission in 1 RTT.

For EC-based reliability we compare two erasure codes:
\begin{enumerate}[leftmargin=*]
\item A simple XOR-based code, in which the \(i\)'th parity block (out of m) is computed as the XOR of all \(k\) data blocks whose indices satisfy \(j \bmod m = i\). This code tolerates the loss of up to one block per each modulo group~\cite{patterson1988case}.
\item An MDS (Maximum Distance Separable) code (e.g., Reed-Solomon), which can recover the data submessage from any \(m\) missing blocks among the total \(m+k\) blocks~\cite{wicker1999reed}.
\end{enumerate}

We study the performance trade-off between these erasure codes. XOR is simpler and easier to optimize for hardware but offers weaker chunk loss tolerance. MDS coding provides stronger protection but is more complex to implement and optimize. In Appendix~\ref{sec:ec-success-derivation}, we derive the success probabilities to decode a data submessage with these schemes. We implement a parallel XOR-based scheme using OpenMP and AVX-512 in $\approx100$ lines of C++ and compare it against Intel’s ISA-L v2.31.1 library~\cite{intel-isal}, which provides an MDS code optimized for Intel CPUs.

\subsubsection{End-to-end SDR testbed}
We validate the ability of SDR SDK to provide its service to the upper reliability layer, a zero-copy message delivery with partial completion semantics, at line rate. We evaluate SDR offloading on nodes of the Israel-1 TOP500 production supercomputer interconnected with 400 Gbit/s RoCEv2 using NVIDIA Spectrum-X~\cite{nvidiaSpectrumXDatasheet}. On each node, we utilize a NVIDIA BlueField-3 SuperNIC connected to a Xeon Platinum 8480 CPU~\cite{nvidiaNVIDIABlueField3Datasheet}. SDR middleware was compiled against DOCA SDK 2.9.0.

\subsection{Deploying SDR at cross-continent scale}\label{sec:two-dcs-case-study}

\begin{figure*}
    \centering
    \includegraphics[width=\textwidth]{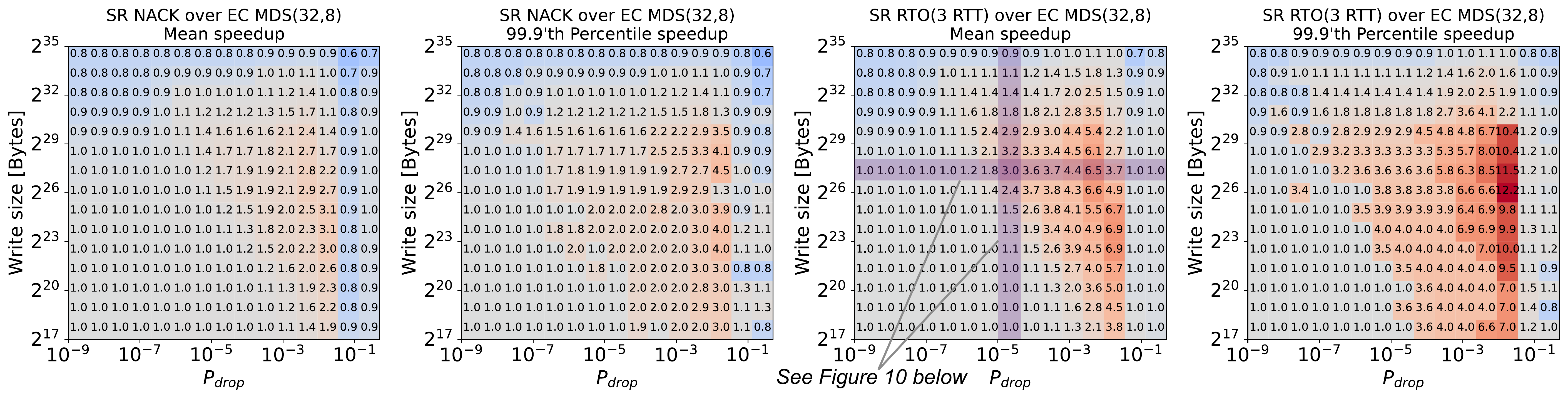}
    \caption{Erasure Coding (EC) improvements (speedup) over Selective Repeat (SR) at 400 Gbit/s and 25 ms RTT.}
    \label{fig:heatmap_percentiles}
\end{figure*}

\begin{figure*}
    \centering
    \includegraphics[width=\textwidth]{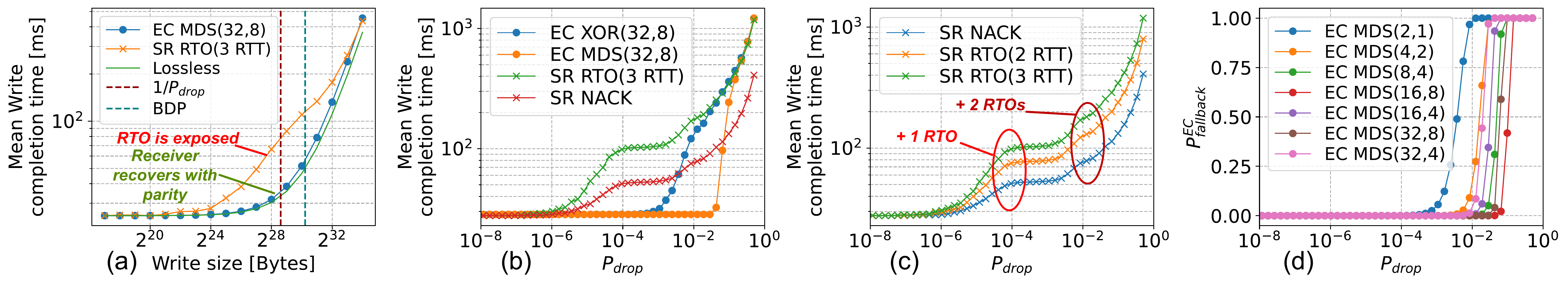}
    \caption{Subplots left to right: (a) variable sized Writes at $P_{drop}=10^{-5}$; (b), (c), (d) 128 MiB Write at various drop rates.
    }
    \label{fig:fec_vs_ec_analysis}
\end{figure*}

\begin{figure}[]
  \centering
  \includegraphics[width=\linewidth]{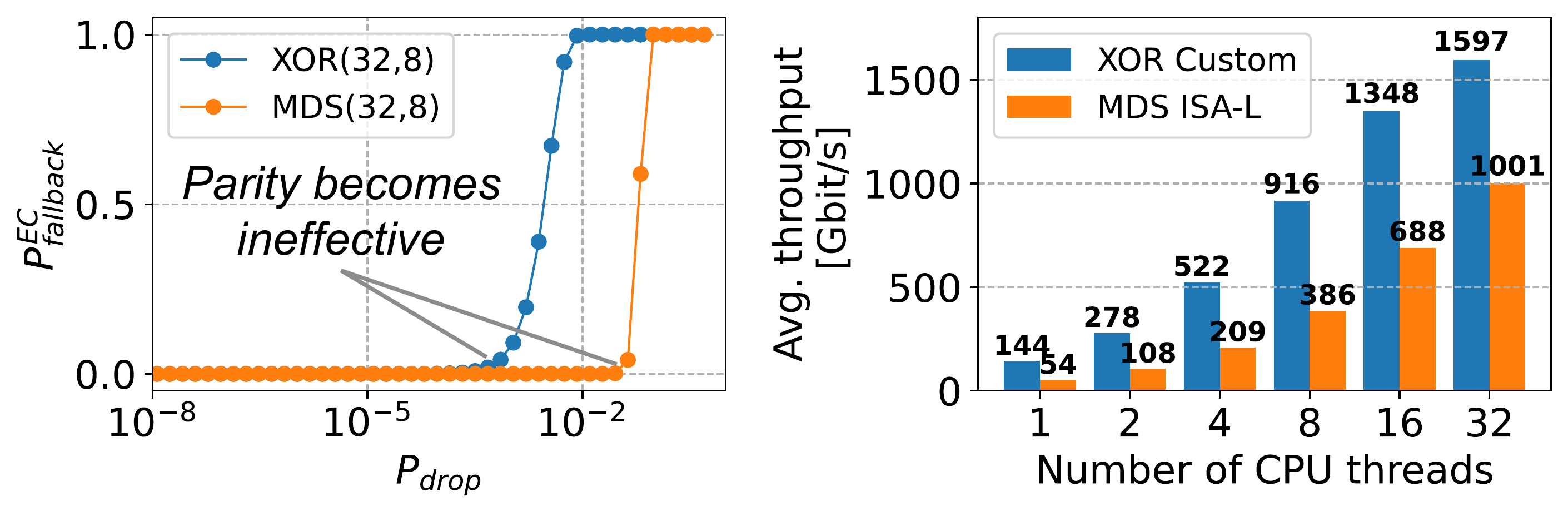}
  \caption{Comparison of MDS EC versus XOR EC. Encoding performance is evaluated on 4 Ghz Intel Xeon Platinum 8580 with 128~MiB buffer, 64~KiB chunk size, $k=32$, $m=8$.}
  \label{fig:ec_performance_xeon}
\end{figure}

\begin{figure}[]
  \centering
  \includegraphics[width=\linewidth]{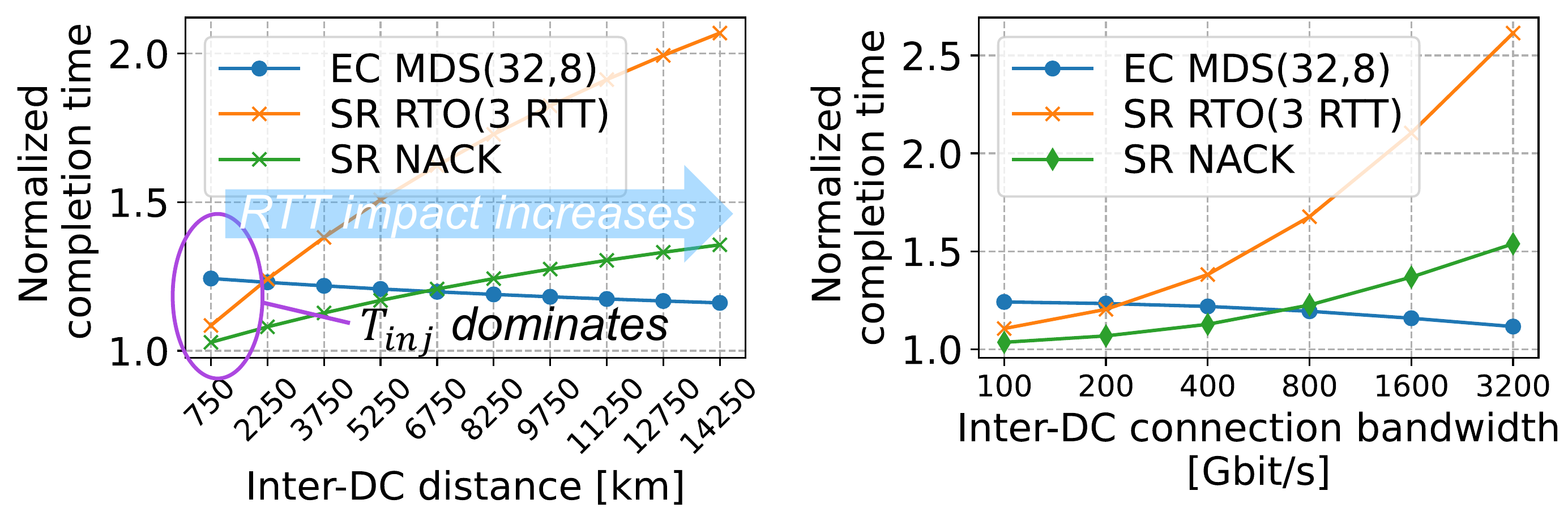}
  \caption{Impact of inter-DC distance and bandwidth on 128 MiB Write completion time. Algorithm times are normalized by a time to perform Write assuming lossless channel.}
  \label{fig:sr_fec_8GiB_RTT_vs_BW}
\end{figure}

\begin{figure}[]
  \centering
  \includegraphics[width=\linewidth]{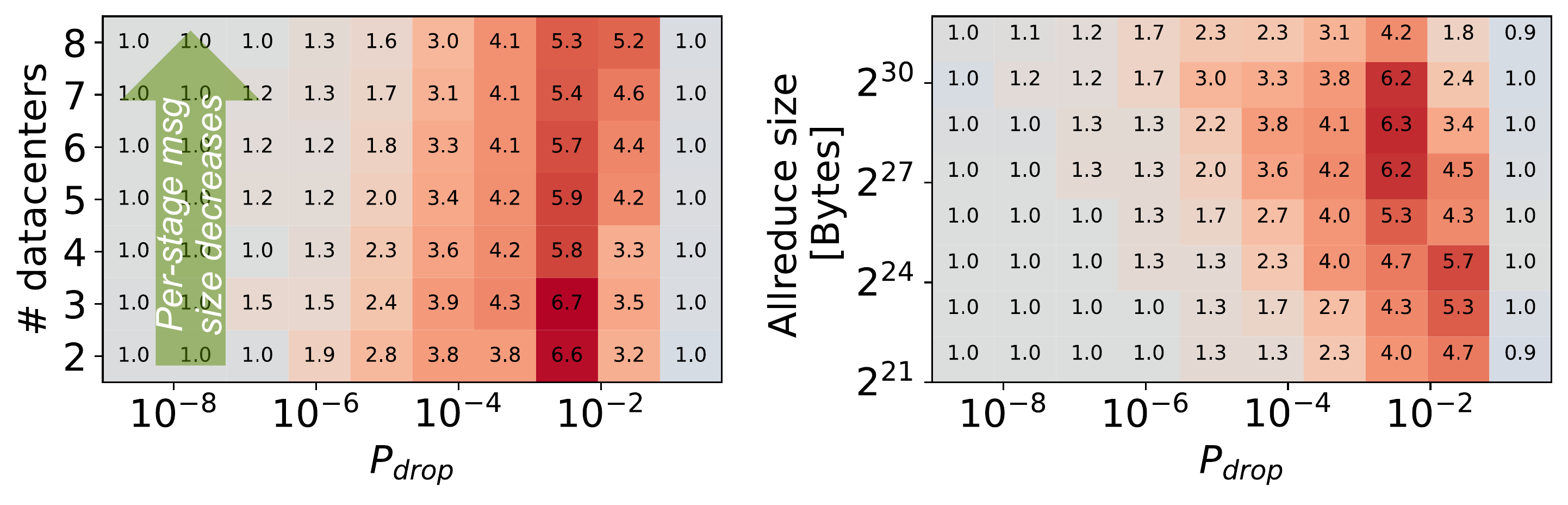}
  \caption{99.9th percentile completion time speedup for inter-datacenter ring Allreduce with MDS EC over SR RTO reliability. Left: 128 MiB buffer size. Right: 4 datacenters.}
  \label{fig:ar-performance}
\end{figure}

\begin{figure}[]
  \centering
  \includegraphics[width=\linewidth]{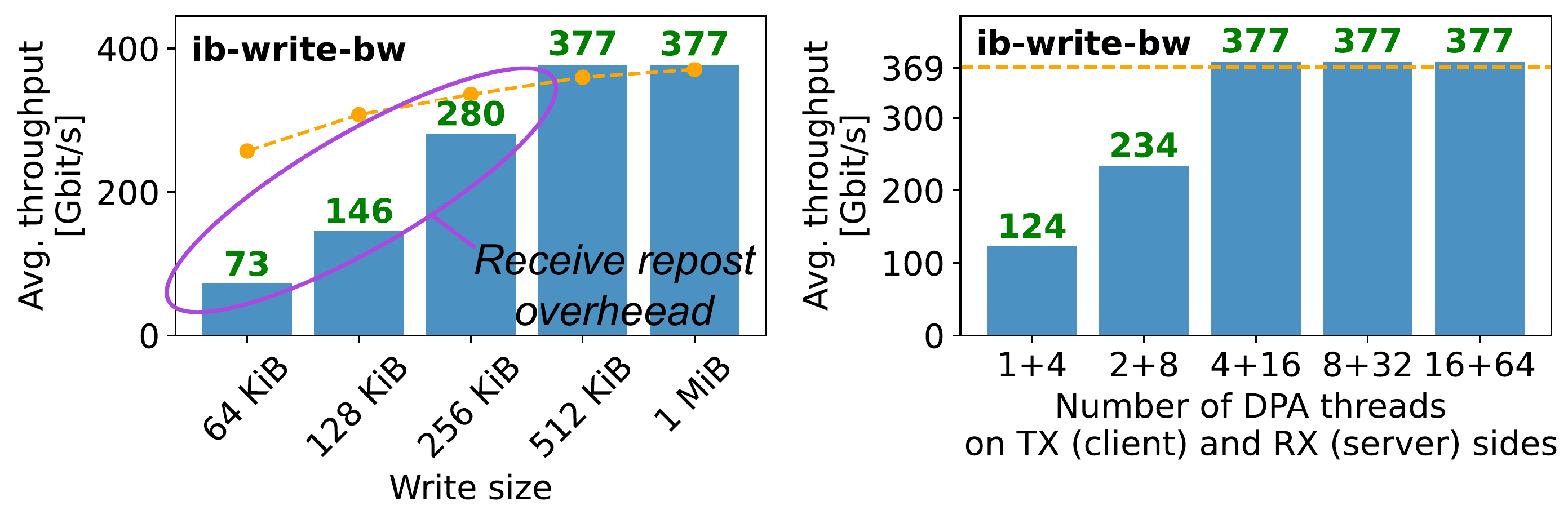}
  \caption{SDR throughput with 16 in-flight Writes and 64~KiB bitmap chunk. Left: throughput vs. message size at 4+16 DPA threads. Right: Thread scaling for 16~MiB messages.}
  \label{fig:urdma_throughput_scaling}
\end{figure}

In our first case study, we analyze SDR performance with a hypothetical 400~Gbit/s, 3750~km cross-continent link between two datacenters (e.g., within the US or Europe). Such a deployment requires the trade-off between cost and channel QoS (i.e., lower drop rates). We examine this trade-off in Figure~\ref{fig:heatmap_percentiles} and provide a detailed analysis of the 128~MiB row and $10^{-5}$ column in Figure~\ref{fig:fec_vs_ec_analysis}.

\subsubsection{Use-case of EC}
In red areas of Figure~\ref{fig:heatmap_percentiles}, EC outperforms SR for messages from $128$~KiB to $1$~GiB within the $10^{-6}$ to $10^{-2}$ drop rate range. For a fixed drop rate in Figure~\ref{fig:fec_vs_ec_analysis}a, a critical message size $1/P$, marks the point beyond which chunk drops become likely. When the Write size nears this point but stays below or comparable to the bandwidth-delay product, SR retransmissions cannot be masked with chunk injection time, leading to slowdowns of up to $6.5\times$ on average and $12.2\times$ at the 99.9th percentile.

The NACK optimization reduces drop detection to 1 RTT, improving SR performance by up to $4\times$ for both average and tail latencies. However, it cannot address the fundamental issue of SR shown in Figure~\ref{fig:fec_vs_ec_analysis}c: RTT-scale penalty per chunk drop.

EC avoids this issue by recovering losses in-place at the receiver. Figure~\ref{fig:fec_vs_ec_analysis}d evaluates various MDS data-parity splits. Lower data-to-parity ratios offer stronger protection at high drop rates with greater bandwidth overhead. We select the $(32,8)$ configuration as the most balanced --- it tolerates drop rates above $10^{-2}$ with no more than 20\% bandwidth inflation.

Efficient EC implementations must hide the encoding overhead. In Figure~\ref{fig:ec_performance_xeon} we assess the compute cost of achieving this by comparing MDS and XOR codes with a $(32,8)$ split. At 400~Gbit/s, XOR encoding can be hidden using 4 CPU cores; MDS needs twice as many. However, XOR trades CPU efficiency for resilience: with a 128~MiB buffer, XOR falls back to SR at $\approx10^{-3}$ drop rate, while MDS remains robust beyond $10^{-2}$.

\subsubsection{When to deploy SR?} 

SR performs best at drop rates below $10^{-6}$ and message sizes above $1$~GiB. An $8$~GiB message, $\approx8\times$ smaller than BDP, is bottlenecked by the injection time, with the final RTT adding little to the total completion time. NACK- and RTO-based SRs hide retransmissions within the injection pipeline, whereas EC introduces a 20\% parity overhead.

Notice that if we were to consider a deployment with a higher RTT or more bandwidth as we do in Figure~\ref{fig:sr_fec_8GiB_RTT_vs_BW}, EC would eventually surpass SR at message size 8~GiB, as retransmissions become more exposed due to increasing BDP.

For small messages in the bottom rows of Figure~\ref{fig:heatmap_percentiles}, SR and EC result in similar completion times. However, due to the compute footprint of EC for path encoding (and decoding in case of drops), SR is preferable. At very high drop rates above 0.1\%, EC is ineffective, as it fails to recover data. As shown in Figure~\ref{fig:fec_vs_ec_analysis}b, MDS coding wastes bandwidth sending parity, ultimately falling back to SR.

\subsection{Inter-datacenter AI collectives}\label{sec:ar-evaluation}

The point-to-point RDMA Write networking primitive, studied in the previous subsection, serves as a building block for collective algorithms. These algorithms are widely used in large-scale parallelized training (e.g., data parallelism)~\cite{nccl, gorentlavenkata2025unified, ben2019demystifying, li2020pytorch, zhao2023pytorch}. Primitives like Allreduce are used for synchronizing model updates across geographically distributed datacenters.

Traditional models, such as LogGP~\cite{alexandrov1995loggp}, assume lossless links between participants. Although valid for intra-datacenter setups (e.g., InfiniBand~\cite{ibspec}, Slingshot~\cite{de2020depth}), this assumption breaks under lossy, high-delay channels, where the choice of reliability scheme becomes paramount, as shown in Figure~\ref{fig:ar-performance}.

In Figure~\ref{fig:ar-performance}, we simulate the performance of the ring Allreduce algorithm across $N$ datacenters with an SDR-based reliability algorithm~\cite{thakur2003mpichcolls}. Tail completion time is strongly affected by the ring schedule, which introduces $2N - 2$ interdependent point-to-point stages. With 4–8 datacenters, messages remain large enough that latency does not dominate, allowing slowdowns from inefficient reliability schemes to accumulate. As a result, the EC scheme, which outperforms SR for drop rates between $10^{-6}$ and $10^{-2}$, also yields gains in multi-stage Allreduce. Across both plots, EC’s speedup over SR grows with the drop rate from $3\times$ to more than $6\times$.

In the Appendix~\ref{sec:ar-lower-bound}, we analytically show that the expected reliability cost in a ring Allreduce is lower bounded by the per-stage cost times the number of stages, explaining the amplified impact on reliability layer efficiency. Our analysis generalizes to other stage-based collective algorithms with schedule dependencies, such as tree algorithms~\cite{sanders2009two}. The SDR framework enables performance engineers to tailor RDMA transport reliability to minimize this cumulative effect in multi-stage protocols.

\subsection{End-to-end performance with BlueField-3}~\label{sec:dpa-eval}

The analysis in previous case studies highlight the importance of flexibility in reliability protocol choice, implementation, and configuration --- a capability that motivated the design of our SDR SDK. In this case study, we subject our offloaded SDR API to a 400 Gbit/s stress test on real hardware.

\subsubsection{SDR client-server performance} We implemented a benchmarking loop on top of SDR middleware API. The benchmark resembles the standard client-server ib\_write\_bw test from the RDMA perftest suite~\cite{rdma-perftest}. For each message, the server emulates a reliability layer by busy polling the completion bitmap. Upon reception of all chunks, server sends an ACK to the client, which runs a timing loop. For each data point, we report average measurements collected from at least 1000 repetitions of the benchmark.

We evaluate SDR's throughput scaling in Figure~\ref{fig:urdma_throughput_scaling}. In typical distributed training workloads, message sizes are often hundreds of megabytes~\cite{zhao2023pytorch, li2020pytorch}. SDR can saturate the link's line rate with much smaller sizes, at the 512~KiB message size SDR needs just 20 out of the 256 available DPA threads to saturate the link. However, for messages smaller than 512~KiB, SDR throughput is behind RC Writes due to software overhead from reposting receive buffers. Each new receive requires message slot reallocation, including memory key table update and bitmap cleanup.

\begin{figure}[]
  \centering
  \includegraphics[width=\linewidth]{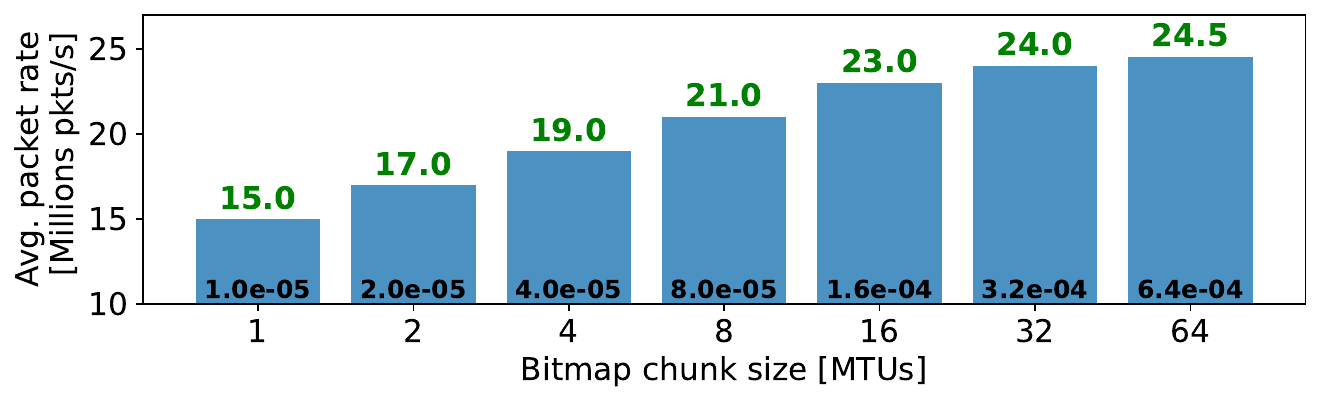}
  \caption{Impact of SDR bitmap chunk size on throughput (shown above bars) and theoretical chunk drop probability~$P_{drop}^{chunk}$ (shown at the bottom of bars) assuming packet (MTU) drop probability $P_{drop}=1e-5$.}
 \label{fig:urdma_bitmap_chunk_size}
\end{figure}

\subsubsection{Impact of the SDR bitmap chunk size}
The SDR bitmap’s variable chunk size lets the reliability layer control how finely it detects network drops. Larger chunks increase the chance of a chunk drop being observed—since a single packet loss causes the entire chunk to be marked as lost: $P_{\text{drop}}^{\text{chunk}} = 1 - (1 - P_{\text{drop}})^N$, where $N$ is the number of packets in the chunk in Figure~\ref{fig:urdma_bitmap_chunk_size}. At the same time, larger chunks also reduce PCIe traffic to the host, as DPA workers update the bitmap only once every $N$ packets.

Interestingly, we observed that 16 receive threads from the best configuration in Figure~\ref{fig:urdma_throughput_scaling} are enough to deliver line rate both at minimum 1-packet chunk of 4096~KiB and maximum 64-packet chunk of 256~KiB. In Figure~\ref{fig:urdma_bitmap_chunk_size} we show further investigation of this phenomena with transport Write size reduced to 64 bytes and bitmap chunk size scaled correspondingly. Reducing chunk size allows us to saturate the link with more packets, while keeping the DPA worker's per-packet load constant, because DPA workers process packet Write completions (not payload!) and their cycle footprint is independent of packet size. We observe that in the worst-case configuration of 1-packet chunks, 16 DPA threads can sustain up to 15~million packets per second, while the theoretical packet rate of 400 Gbit/s link at 4~KiB MTU is 11.6~million. This experiment shows that SDR API offloading enables control of the drop rate at the upper reliability layer without compromising performance.

\subsubsection{SDR performance with Tbit/s links}

\begin{figure}[]
  \centering
  \includegraphics[width=\linewidth]{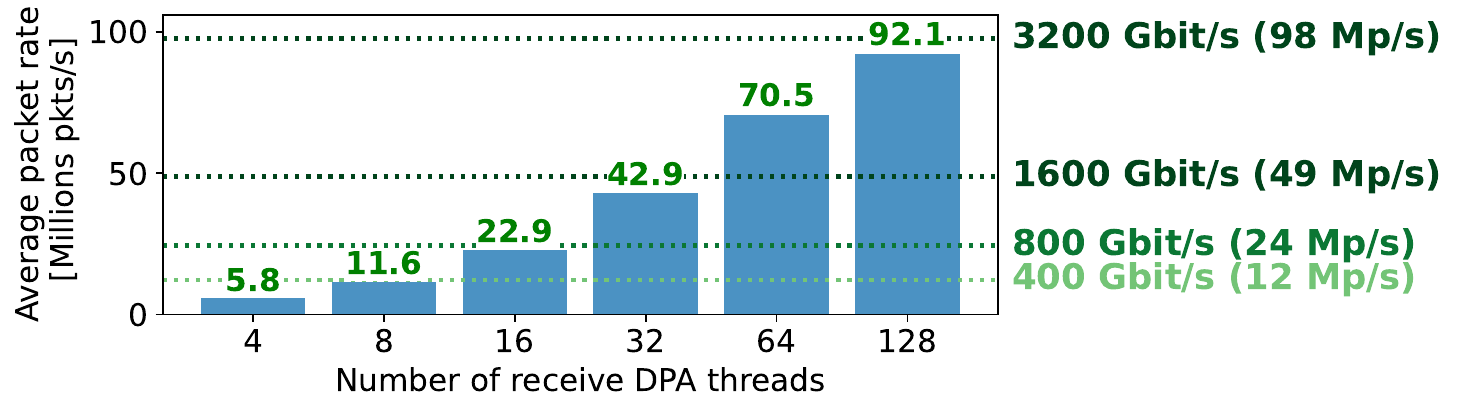}
  \caption{SDR packet rate scaling versus the number of DPA threads used for receive side offloading.}
  \label{fig:urdma_msg_rate_scaling}
\end{figure}

Finally, in Figure~\ref{fig:urdma_msg_rate_scaling} we examine SDR's ability to serve traffic at packet rates expected with next-generation Tbit/s link bandwidths at 4 KiB MTUs and 64 KiB chunks. To maximize packet rate load at the receiver, we use the methodology from the previous example and use 4 CPU threads to generate 64 byte packets on the client side. DPA-based offloading scales nearly linearly across 4 to 32 threads. At 32 threads (1/8 of DPA capacity), SDR reaches packet rates near 1.6 Tbit/s. Scaling to 128 threads brings throughput close to 3.2 Tbit/s. This demonstrates that SDR’s multi-channel backend, combined with DPA offloading, decouples per-packet progress from upper-layer reliability and ensures scalability for next-generation links.

\section{Related Work}

IRN~\cite{mittal2018irn} and SRNIC~\cite{wang2023srnic} address design limitations of commodity RDMA NICs. Like SDR, they use per-connection bitmaps for retransmission-based reliability; however, these bitmaps are internal to the FPGA and hidden from users, limiting experimentation with alternative schemes and compatibility with commodity NICs. Their wire protocols are also incompatible with commodity NICs.

Flor~\cite{li2023flor} enhances Go-back-N reliability in ConnectX-4/5 RC transport by supporting selective retransmission an top of UC. Unlike SDR’s partial message completion, it lacks a unified abstraction for general transport-layer reliability and does not support offloading. Khalilov et al.~\cite{khalilov2024network} use a software bitmap for multicast-based collectives. Their offload-oriented design also supports UC but does not generalize to arbitrary traffic, unlike SDR.

Two decades ago, Lundqvist and Karlsson~\cite{lundqvist2004tcp} showed that end-to-end FEC can significantly boost TCP Reno, SACK, and Tahoe throughput for Internet traffic. Maelstrom~\cite{balakrishnan2008maelstrom} introduces a proxy-based design that applies FEC to UDP traffic at the datacenter edge.

More recent works, Cloudburst~\cite{zeng2022cutting} and LoWAR~\cite{zuo2024lowar} apply FEC to the datacenter transport layer. Cloudburst distributes coded packets across parallel paths for early recovery, while LoWAR targets long-haul links. Both outperform retransmission protocols but lack scalability analysis for next-generation links, limit users to FEC-based reliability and work on top of custom transport.

\section{Conclusion}

We presented SDR, a novel software-defined RDMA stack that enables custom reliability algorithms for long-haul RDMA across datacenters. SDR introduces partial message completion via a bitmap API, empowering developers to tailor reliability strategies --- such as Selective Repeat and Erasure Coding --- to specific network conditions without sacrificing RDMA’s zero-copy performance. By offloading packet-processing logic to the NIC, SDR achieves full line rate performance on current generation hardware and supports packet rates of next-generation Tbit/s links. SDR offers immediate, deployable improvements over existing NIC solutions, unlocking optimized inter-datacenter GPU training communication.

\begin{acks}
We thank CSCS and Jérôme Tissières for providing the infrastructure used to perform some of the experiments.
This work is supported by the following grant agreements: SwissTwins (funded by the swiss State Secreteriat for Education, Research and Innnovation), ERC PSAP (grant agreement No 101002047), WeatherGenerator (grant agreement No 101187947).
\end{acks}

\bibliographystyle{ACM-Reference-Format}
\bibliography{bibliography}

\appendix

\section{Analytical solution for message completion time expectation}~\label{sec:fct-derivation}

In the model for Selective Repeat (SR) presented in Section~\ref{sec:fct-model}, we define the message completion time as \(\max X_i\). In this appendix, we provide an analytical derivation for \(\max X_i\).

The event \(\{X_i \ge q\}\) is equivalent to
\[
t_{\text{start}}(i) + O\cdot \,\bigl(Y_i - 1\bigr) \ge q.
\]
Rearranging to isolate the integer-valued \(Y_i\), we obtain
\[
Y_i \ge \left\lceil 1 + \frac{q - t_{\text{start}}(i)}{O} \right\rceil.
\]
Since \(Y_i \sim \operatorname{Geom}(1 - P_{drop})\), we have
\[
P(Y_i \ge k) = P_{drop}^{\,k - 1}, \quad k \ge 1.
\]
Thus,
\[
P\left(Y_i \ge \left\lceil 1 + \frac{q - t_{\text{start}}(i)}{O} \right\rceil\right)
= P_{drop}^{\,\left\lceil 1 + \frac{q - t_{\text{start}}(i)}{O} \right\rceil - 1}
= P_{drop}^{\,\left\lceil \frac{q - t_{\text{start}}(i)}{O} \right\rceil}.
\]
Therefore, the tail probability for \(X_i\) is
\[
P(X_i \ge q) =
\begin{cases}
1, & q \le t_{\text{start}}(i), \\[1mm]
P_{drop}^{\,\left\lceil \frac{q - t_{\text{start}}(i)}{O} \right\rceil}, & q > t_{\text{start}}(i).
\end{cases}
\]

Assuming independence among chunks, the tail probability for \(\max X_i\) is
\[
P(\max X_i \ge q) = 1 - \prod_{i=1}^{M} P\bigl(X_i < q\bigr)
= 1 - \prod_{i=1}^{M} \left[1 - P(X_i \ge q)\right].
\]

Using the tail-sum formula~\cite{bertsekas2008introduction}, the expectation is given by
\[
\mathbb{E}[\max X_i] = \sum_{q=1}^{\infty} P(\max X_i \ge q).
\]

Substituting the expression for \(P(\max X_i \ge q)\) and accounting for the round-trip time to receive the final ACK, the expected message completion time is
\[
\mathbb{E}[T_{SR}(M)] = \sum_{q=1}^{\infty} \left\{ 1 - \prod_{i=1}^{M} \left[ 1 - P(X_i \ge q) \right] \right\} + RTT.
\]

\section{Probability of\\ successful decoding for MDS and XOR EC}~\label{sec:ec-success-derivation}
In Section~\ref{sec:experimental-setup}, we consider two submessage EC schemes: MDS and XOR. In this appendix, given a chunk drop probability \(P_{drop}\), a data submessage of \(k\) chunks, and a parity submessage of \(m\) chunks, we derive the probability of successful data submessage recovery for these schemes.

\subsubsection{MDS EC}
Let's denote \(X\) as the number of chunk drops within the data and parity submessages. The probability of successful recovery of a data submessage using MDS erasure coding is:
\[
P_{EC(k,m)}^{MDS} = P(X \leq m) = \sum_{i=0}^{m} \binom{k+m}{i} P_{drop}^{i} (1 - P_{drop})^{k+m - i}.
\]

\subsubsection{XOR EC} 

Let's denote the number of chunks within an XOR modulo group as \(n = k/m + 1\). The probability of successful recovery of a data submessage using XOR erasure coding is:

\[
P_{EC(k,m)}^{XOR} = [(1 - P_{drop})^{n} + \binom{n}{1} P_{drop}^{1} (1 - P_{drop})^{n - 1}]^{m} =
\]
\[
= [(1 - P_{drop})^{n} + n P_{drop} (1 - P_{drop})^{n - 1}]^{m}.
\]

\section{Lower Bound on \\ Expected Allreduce Completion Time}~\label{sec:ar-lower-bound}

In Section~\ref{sec:ar-evaluation}, we empirically studied the performance of an inter-datacenter Allreduce collective. Let's assume that this collective utilizes the reliable Write as a peer-to-peer (P2P) transport. The message completion time for Write was derived in Section~\ref{sec:reliability-layer}. Here, we derive a theoretical lower bound on the expected completion time for Allreduce operation, highlighting the accumulated impact of reliability costs.

Consider $N$ datacenters participating in the Allreduce. The ring Allreduce algorithm involves $2N-2$ sequential rounds of P2P communication steps~\cite{thakur2003mpichcolls}. Let $T(i, r)$ denote the finish time of round $r$ at datacenter $i$, where $i \in \{0, \dots, N-1\}$ (indices are often considered modulo $N$) and $r \in \{0, \dots, 2N-2\}$. We assume all datacenters start synchronously at time $t=0$, thus the initial condition is $T(i, 0) = 0$ for all $i$.

The completion time of round $r$ at datacenter $i$, $T(i,r)$, depends on when both datacenter $i$ and its predecessor $i-1$ (modulo $N$) finished round $r-1$. The task for round $r$ at node $i$ can only commence after $\max(T(i-1, r-1), T(i, r-1))$. Let $t(i, r-1)$ denote the duration of the P2P communication and processing step associated with completing round $r$ at node $i$, which involves receiving data sent by node $i-1$ after its round $r-1$. The finish time recurrence is modeled as:
\begin{equation} \label{eq:recurrence_T}
T(i, r) = \max(T(i-1, r-1), T(i, r-1)) + t(i, r-1) \quad \text{for } r \ge 1
\end{equation}
The duration $t(i, k)$ accounts for both baseline P2P communication time and potential delays due to reliability mechanisms. We model it as:
\begin{equation} \label{eq:t_model}
t(i, k) = C + X(i, k)
\end{equation}
where $C$ is a constant cost representing the transmission time under ideal (e.g., lossless) network conditions, and $X(i, k)$ is a non-negative random variable representing the additional delay incurred due to the reliability protocol (e.g., acknowledgments, retransmissions) for the communication step indexed by $k$.

We aim to find a lower bound on the expected completion time of the Allreduce, which corresponds to the expected finish time of the last round, $\mathbb{E}[T(i, 2N-2)]$. Taking the expectation of Equation~\eqref{eq:recurrence_T}:
\[
\mathbb{E}[T(i, r)] = \mathbb{E}[\max(T(i-1, r-1), T(i, r-1)) + t(i, r-1)]
\]
Assuming the random duration $t(i, r-1)$ is independent of the past finish times $T(i-1, r-1)$ and $T(i, r-1)$ (a common simplifying assumption for deriving bounds), we can use the linearity of expectation:
\[
\mathbb{E}[T(i, r)] = \mathbb{E}[\max(T(i-1, r-1), T(i, r-1))] + \mathbb{E}[t(i, r-1)]
\]
By Jensen's inequality, since the $\max(\cdot, \cdot)$ function is convex, we have $\mathbb{E}[\max(A, B)] \geq \max(\mathbb{E}[A], \mathbb{E}[B])$. Applying this yields:
\begin{equation} \label{eq:jensen_step}
\mathbb{E}[T(i, r)] \geq \max(\mathbb{E}[T(i-1, r-1)], \mathbb{E}[T(i, r-1)]) + \mathbb{E}[t(i, r-1)]
\end{equation}
Let $\mu_X = \mathbb{E}[X(i, k)]$ be the expected reliability cost per step. We assume $\mu_X$ is constant across all datacenters $i$ and steps $k$ for simplicity. Consequently, the expected duration of a step is $\mathbb{E}[t(i, k)] = C + \mu_X$.

Furthermore, we assume statistical symmetry across datacenters. This implies that the expected finish time for any round $r$ is independent of the specific datacenter index $i$. Let $E_r = \mathbb{E}[T(i, r)]$ denote this common expected finish time for round $r$. Substituting this notation and the expected step duration into Inequality~\eqref{eq:jensen_step}:
\[
E_r \geq \max(E_{r-1}, E_{r-1}) + (C + \mu_X)
\]
which simplifies to the linear recurrence inequality:
\begin{equation} \label{eq:recurrence_E}
E_r \geq E_{r-1} + C + \mu_X \quad \text{for } r = 1, 2, \dots, 2N-2
\end{equation}
We can unroll this recurrence starting from the initial condition $E_0 = \mathbb{E}[T(i, 0)] = 0$:
Therefore, the expected completion time of the ring Allreduce operation using this reliable P2P model is lower bounded by:
\begin{equation} \label{eq:lower_bound}
E_{2N-2} \geq (2N-2)(C + \mu_X)
\end{equation}
This result clearly demonstrates the accumulation of delay: the expected cost of reliability per step, $\mu_X$, is multiplied by the total number of sequential steps ($2N-2$) in the ring Allreduce algorithm. This highlights how reliability mechanisms can significantly impact the overall performance, especially in environments where $\mu_X$ is non-negligible relative to $C$.

\end{document}